\documentclass[twocolumn,aps,prx,superscriptaddress,longbibliography]{revtex4-2}
\usepackage[colorlinks,bookmarks=false,citecolor=blue,linkcolor=red,urlcolor=blue]{hyperref}
\usepackage{verbatim}
\usepackage{color}
\usepackage{amsmath,amssymb,bm,braket,float,mathtools}
\usepackage{graphicx,xfrac,appendix}
\usepackage[english]{babel}

\begin{document}	
\title{Stable real-energy spectral dynamics with topological transitions and non-Hermitian many-body localization}
\author{Shujie Cheng}
\thanks{chengsj@zjnu.edu.cn}
\thanks{2818917376@qq.com}
\affiliation{Xingzhi College, Zhejiang Normal University, Lanxi 321100, China}
\affiliation{Department of Physics, Zhejiang Normal University, Jinhua 321004, China}
\author{Xixi Feng}
\affiliation{Department of Physics, Zhejiang Normal University, Jinhua 321004, China}
\author{Wen Chen}
\affiliation{Beijing Computational Science Research Center, Beijing 100193, China}
\author{Niaz Ali Khan}
\affiliation{Department of Physics, Zhejiang Normal University, Jinhua 321004, China}
\author{Gao Xianlong}
\thanks{Corresponding author: gaoxl@zjnu.edu.cn}
\affiliation{Department of Physics, Zhejiang Normal University, Jinhua 321004, China}


\begin{abstract}
In this work, the interplay between non-Hermiticity, quasi-disorder, and repulsive interaction is studied for hard-core bosons
confined in a one-dimensional optical lattice, where non-Hermiticity is induced by the non-reciprocal
hoppings and the on-site gain and loss breaking the time-reversal symmetry. Although the energy spectra of the static
system are fully complex, with the evolution of the initial state, the real part of the expectation value of the Hamiltonian under
the time-evolved wave function changes stably.  By means of the entanglement entropy and its dynamical evolution,
as well as the inverse participation ratio, the many-body localization (MBL) is found to play the key role in the stability of
the dynamical behavior of the real part of the expectation value, independent of whether the spectrum of the static
Hamiltonian is real or complex.  In the delocalization phase, the dynamical evolution of the real part of the expectation value is unstable.
Meanwhile, the nearest-neighbor level spacings statistics shows the MBL transition accompanied by the transition from
the Ginibre distribution to the complex Poisson distribution, different from the one in the time-reversal invariant system.
In addition, the dynamical stability of the real part of the energy and the MBL transition can be characterized by the winding number,
indicating that the MBL transition and the topological transition occur simultaneously, and the realization of the Hamiltonian is
discussed.

\end{abstract}
\maketitle

\section{Introduction}\label{S1}

The eigenstate thermalization hypothesis \cite{ETH-1,ETH-2,ETH-3,ETH-4,ETH-5,Wigner-Dyson-ETH}
is a widely accepted concept that explains the thermalization of the eigenstates. It refers to that the eigenstates
of the Hamiltonian of an isolated system will themselves exhibit the properties of heating, resulting in these thermalized
eigenstates being ergodic. However, the many-body localization (MBL) violates this hypothesis,
making it an active field in condensed physics over the past decade. In the thermalization phase, spectral statistics
indicates that the energy spectrum follows the Wigner-Dyson distribution while in the MBL phase,
the Poisson distribution \cite{Wigner-Dyson-ETH,Wigner-Dyson}.
The difference in statistical distributions before and after the MBL transition attracts wide
interest in the study of spectral properties of
Gaussian and Poisson ensembles \cite{Hermite_statistic_1,Hermite_statistic_2,Hermite_statistic_3,Hermite_statistic_4,Hermite_statistic_5,Hermite_statistic_6}.

In fact, spectral statistics is one of the tools to understand the thermalization to MBL from the energy perspective.
The MBL transition can be characterized by the many-body eigenstate as well. The half-chain entanglement entropy
is obtained from the subspace decomposition of eigenstates. It has been shown that
for the thermal phase where eigenstates are delocalized, the entanglement entropy follows the volume law, while in the MBL
phase, the area law \cite{entropy-1,entropy-2,entropy-3,entropy-4,entropy-5,entropy-6,entropy-7,entropy-8,
entropy-9,entropy-10,entropy-11,entropy-12,entropy-13,entropy-14,entropy-15}. Additionally, the MBL transition
is observed through diagonal entropy \cite{diagonal-entropy-1,diagonal-entropy-2,entropy-diagonal-entropy-fisher-info},
local parameter, which weighs the stability of the eigenstates under local perturbations,
and grows with system size in the delocalized phase and decreases in the MBL phase \cite{local-parameter} , many-body inverse participation ratio (IPR) \cite{IPR}, or normalized participation ratio \cite{NPR},
quantum Fisher information \cite{entropy-diagonal-entropy-fisher-info,ion-fisher-info-1,ion-fisher-info-2,fisher-info-1,fisher-info-2},
and particle imbalance \cite{atom-particle-imb-1,atom-particle-imb-2,atom-particle-imb-3,atom-particle-imb-4,atom-particle-imb-5,
atom-particle-imb-6,atom-particle-imb-7,particle-imb-1,particle-imb-2,particle-imb-3,particle-imb-4,particle-imb-5,particle-imb-6,Suthar_0}.
Experimentally, the MBL has been realized in various controllable platforms, such as
the ultracold atomic systems \cite{atom-particle-imb-1,atom-particle-imb-2,atom-particle-imb-3,atom-particle-imb-4,atom-particle-imb-5,
atom-particle-imb-6,atom-particle-imb-7,atom}, trapped ions \cite{ion-fisher-info-1,ion-fisher-info-2},
superconducting processors \cite{processor_1,processor_2,processor_3,processor_4,processor_5,processor_6},
nuclear spins \cite{nuclear-spin}, and solid material \cite{phonon-MBL}.

Similarly to the Hermitian systems, the level statistics can be applied to non-Hermitian systems as well.
Differently, the non-Hermiticity leads to a new level statistics distribution, namely the Ginibre distribution \cite{Ginibre-origion}.
Due to the emergence of non-Hermiticity, the originally Hermitian 10-fold topological classifications \cite{symmetry-10-origion,symmetry-10-2,symmetry-10-3}
have been generalized to the 38-fold topological classifications \cite{symmetry-38-1,symmetry-38-2,symmetry-38-3}. Accordingly,
the Ginibre distribution is widely studied in the open quantum systems with various
symmetries \cite{Ginibre-MBL-entropy,Ginibre-symmetry,Ginibre-1,Ginibre-2,Ginibre-3,Ginibre-4,Ginibre-5,Ginibre-6,Ginibre-7,Ginibre-8,Ginibre-9}.
However, it is constrained by the symmetry of non-Hermitian systems. In general,
the level statistics in the systems with transposition symmetry (Classes ${\rm AI}^{\dag}$ and ${\rm AII}^{\dag}$ ) will
deviate from the Ginibre distribution \cite{Ginibre-MBL-entropy,Ginibre-symmetry,Singular_Value_Statistics_kawabata}.

Recently, non-Hermitian spectral statistics has been employed to analyze the MBL transition. In a class of
non-Hermitian many-body disordered or quasi-disordered systems with a complex-real transition
in energy,  level statistics in the delocalized phase (with complex energies) presents the Ginibre distribution, while it presents the
real Poisson distribution in the MBL phase (with real energies) \cite{Ginibre-MBL-entropy,NH_MBL_3,NH_MBL_4,Suthar_1}. Just as in the Hermitian case,
the half-chain entanglement entropy can still be used to extract the transition point and the scaling
exponent of the MBL transition. Additionally, for the delocalized states, the entanglement entropy follows
the volume law, while for the MBL states, it follows the area law \cite{Ginibre-MBL-entropy,NH_MBL_2,NH_MBL_3,NH_MBL_4,Suthar_1}.
It is worth noting that, for the delocalized phase, i.e., the complex energy regime, the real part of the energy
is unstable during the time evolution, while for the MBL phase, i.e., the real energy regime, the dynamical
process is stable \cite{Ginibre-MBL-entropy,NH_MBL_3}. This raises the question of whether MBL or real energy
plays the key role in maintaining the stable dynamical behavior, which is not addressed in the previous works \cite{Ginibre-MBL-entropy,NH_MBL_3}. In this paper, we attempt to study a non-Hermitian many-body system with a
fully complex energy spectrum and with MBL to answer this question and give a clear understanding on the stable dynamical behavior.
We note that for the non-Hermitian many-body systems with complex-real energy transition,
the winding number \cite{NH_winding_1,NH_winding_2} can be used to characterize the MBL transition
(where non-zero winding number corresponds to the delocalized phase whereas the zero winding number
corresponds to the MBL phase) \cite{NH_MBL_3,NH_MBL_4}. However, the topological transition in a complex spectral case is not uncovered \cite{NH_MBL_4}. Whether the topological transition occurs in complex spectral systems, and if the topological transition exists, whether it follows the MBL transition, demand the answer. Besides we will discuss the relationship between the topological and MBL transitions.
Some researches showed that the exceptional point \cite{Heiss_1990,Heiss_2005,Heiss_2012,PRE2018} leads to the nontrivial topology \cite{Kawabata_PRB,Kawabata_PRL,PRR_Luitz}. We will go into the topological origin
of the complex energy spectrum system and investigate the connection between topological transitions and
exceptional points.

The paper is organized as follows. Section \ref{S2} introduces the model and Hamiltonian. Section \ref{S3} analyzes and discusses the
properties of the energy spectrum, including the complex energy spectrum and  the energy evolution.
Section \ref{S4} analyzes and discusses the MBL transition, including the spectrum statistic, half-chain entanglement entropy, entropy evolution,
scaling exponent of entropy, the many-body IPR, and the time-dependent density imbalance.
Section \ref{S5} studies the topological transition and introduces how to realize the Hamiltonian.
Section \ref{S6} presents the summary.

\section{Model and Hamiltonian}\label{S2}
 Here, we study a non-Hermitian one-dimensional hard-core bosonic system and the Hamiltonian is
 \begin{equation}\label{Ham}
 \hat{H}=\sum^{L}_{j=1}\left[-J(e^{-g}\hat{c}^{\dag}_{j+1}\hat{c}_{j}+e^{g}\hat{c}^{\dag}_{j}\hat{c}_{j+1})+U\hat{n}_{j}\hat{n}_{j+1}+V_j\hat{n}_{j}\right],
 \end{equation}
where $\hat{c}_{j}$ ($\hat{c}^{\dag}_{j}$) is the particle annihilation (creation) operator, $\hat{n}_{j}=\hat{c}^{\dag}_{j}\hat{c}_{j}$
is the particle-number operator. Mismatched hopping strengths $-Je^{-g}$ and $-Je^{g}$ form
nonreciprocal hoppings with $J$ the unit of energy and $g$ the dimensionless parameter.
$U$ is the repulsive interaction between nearest-neighbor sites. Here, the quasi-disordered on-site potential
$V_{j}$ with odd-site gain ($+i\gamma$) and even-site loss ($-i\gamma$) is considered, i.e., $V_{j}=V\cos(2\pi\alpha j+\varphi)-i\gamma(-1)^{j}$,
where $V$ is the strength of the quasi-disorder potential, $\varphi$ is a random phase and $\alpha$ is the incommensurate parameter.

The non-Hermiticity of this system is controlled by the parameters $g$ and $\gamma$ together.  When $\gamma=0$, if we perform a
transformation, i.e., $\hat{c}_{j}\rightarrow e^{-gj}\hat{c}_{j}$ and $\hat{c}^{\dag}_{j}\rightarrow e^{gj}\hat{c}^{\dag}_{j}$,
the Hamiltonian presented in Eq. (\ref{Ham}) can go back to a Hermitian many-body interacting model \cite{Huse_MBL_AA}, in which the authors
give the numerical estimates for the MBL transition point. For non-zero $\gamma$, the model in Eq. (\ref{Ham}) breaks the time-reversal symmetry.
In the following, we take $U=2J$ and $g=0.5$ as an example to reveal the MBL property and its relationship with topological transition
in the absence of time-reversal symmetry. The strength of the quasi-disorder potential $V$ is chosen as the control
parameter for the underlying transitions. The incommensurate parameter is chosen at $\alpha=(\sqrt{5}-1)/2$ and the periodic
boundary condition is considered in the following analyses. All the calculations are performed in the subspace with particles $N=L/2$.

\begin{figure}[htp]
	\centering
	\includegraphics[width=0.5\textwidth]{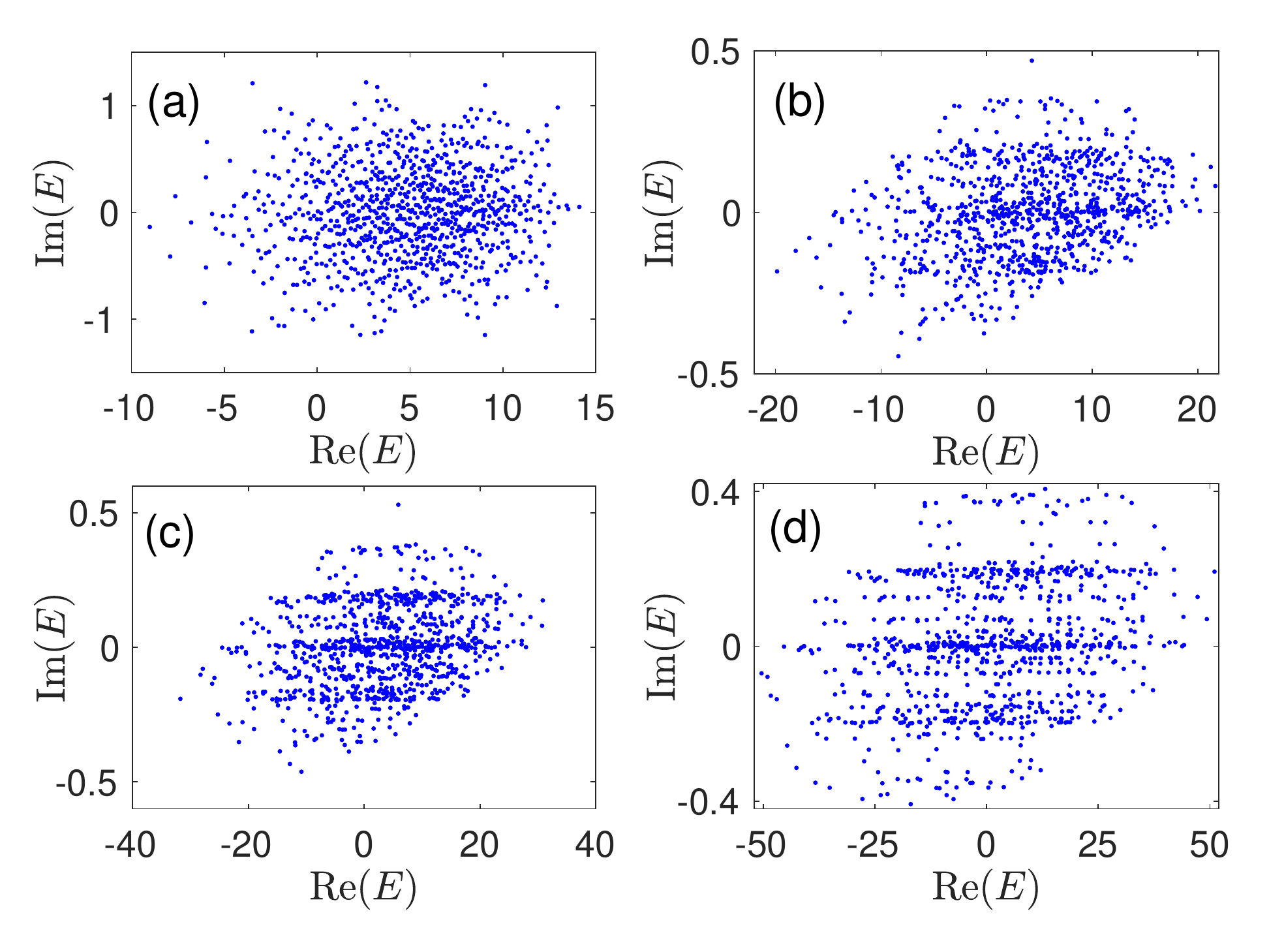}
	\caption{Energy spectra in the complex plane. (a) $V=2J$. (b) $V=5J$.
	(c) $V=8J$. (d) $V=14J$. Other parameters are $L=12$, $\gamma=0.1J$, and $\varphi=\pi/4$.} \label{f1}
\end{figure}

\section{Energy spectrum properties}\label{S3}
At first, the energy spectrum characteristic is discussed. In the presence of time-reversal symmetry, i.e., $\gamma=0$,
we have known that the energy spectra present a complex-to-real transition \cite{NH_MBL_3}. However,
in the absence of time-reversal symmetry, i.e., $\gamma\neq 0$, we find that there is no complex-to-real transition. Taking
$\gamma=0.1J$, $\varphi=\pi/4$, and various $V$, the resulting spectra are plotted in in Figs.~\ref{f1}(a)-\ref{f1}(d).
Intuitively, as the increase of the strength of quasi-disorder potential, there is always nonvanishing imaginary part of
the energy spectrum in the complex plane. It means that the time-reversal symmetry is essential for the system to
have a static real energy spectrum. Meanwhile, we note that for the systems with time-reversal symmetry, the
energy spectra characterized by the evolution of the real part of the energy is dynamically stable in the real energy regime \cite{Ginibre-MBL-entropy,NH_MBL_3}.
It raises a question whether the real energy (or the time-reversal symmetry) is essential in
maintaining the stably dynamical behaviors. Next, we discuss the dynamical stability of the
energy spectra which is characterized by the evolution of the real part of the energy
$E^{R}(t)={\rm Re}[\overline{\langle \psi(t)|\hat{H}|\psi(t)\rangle}]$, where $\ket{\psi(t)}$ is the
wave function evolving from the initial state to time $t$ \cite{Ginibre-MBL-entropy}.
and  $\overline {~~\cdot~~}$ denotes the ensemble average. We emphasize that the disorder configuration
of the random phase $\varphi$ is determined at $t=0$, and do not change at $t>0$.
 $\ket{\psi(t)}=e^{-i\hat{H}t}\ket{\psi(t=0)}/\sqrt{\mathcal{N}}$
describes a quantum trajectory without any quantum jumps, which is microscopically justified in the context of continuously measured
systems \cite{Ginibre-MBL-entropy,time_evolution}. Here, the initial state $\ket{\psi(t=0)}$ is chosen as $\ket{101010\cdots}$ and
$\mathcal{N}$ is normalization coefficient, defined as $\mathcal{N}=\langle \psi(t) | e^{iH^{\dag}t}e^{-iHt} |\psi(t) \rangle$.
According to the Baker-Campbell-Hasusdorff expansion, we know that $e^{iH^{\dag}t}e^{-iHt}$ contains the difference term $H^{\dag}-H$
and the commutation term $\left[H^{\dag}, H\right]$. Due to the non-Hermiticity of $H$, we have $\left(H^{\dag}-H \right)\neq 0$
and $\left[H^{\dag}, H\right]\neq 0$. Therefore, $\ket{\psi(t)}$ and $\mathcal{N}$ will change during the time evolution,
and may lead to the time dependence of $E^{R}(t)$.

\begin{figure}[htp]
	\centering
	\includegraphics[width=0.5\textwidth]{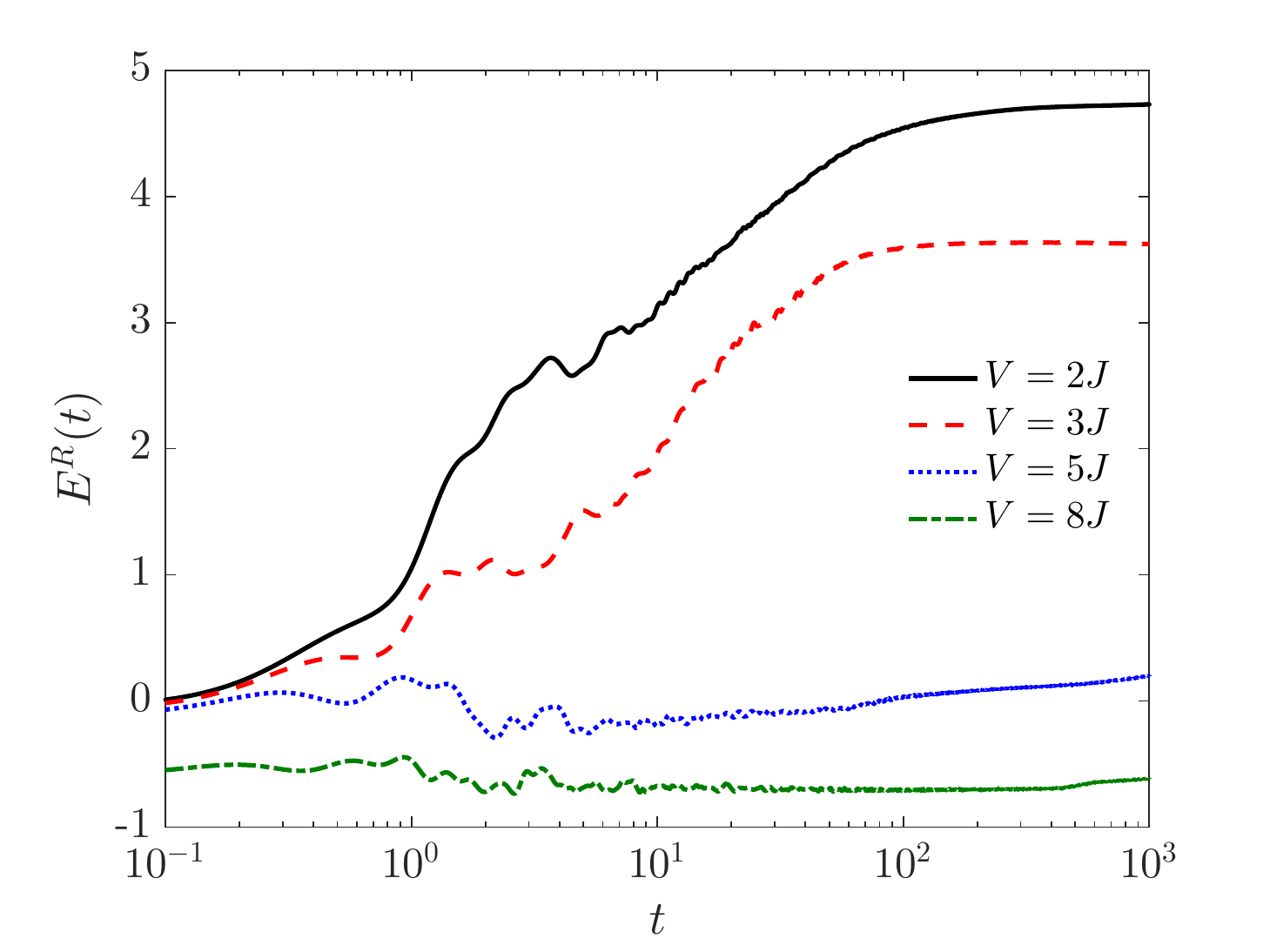}
	\caption{Time evolution of $E^{R}(t)$ for $V=2J$, $3J$, $5J$, and $8J$. The system size is $L=12$ and the
	initial state is taken as $\ket{\psi(t=0)}=\ket{101010\cdots}$. We take $\gamma=0.1J$ and $512$ samples in the calculation.} \label{f2}
\end{figure}

Taking $L=12$, $\gamma=0.1J$, and $\varphi=\pi/4$, we plot the time evolution of $E^{R}(t)$ $t$ under various $V$
in Fig.~\ref{f2}. We take $256$ samples in the calculations. We can see that when the potential strengths are relatively weak,
such as $V=2J$ (black curve) and $V=3J$ (red curve), $E^{R}(t)$ are quite unstable and evidently deviates from the initial state
during the dynamical process. However, It seems that there exists a phase transition which leads to the different dynamical
process of $E^{R}(t)$. For $V=5J$ (blue curve) and $V=8J$ (green curve), the dynamical processes are relatively stable, because during the time
evolution, $E^{R}(t)$ only slightly deviates from the initial state. Even with a longer evolution time, such as $t \in ( 10^{2},10^{}3)$, the dynamical
evolutions of $E^{R}(t)$ under $V=5J$ and $V=8J$ are still stable. Recalling that in the presence of time-reversal symmetry,
the energy spectra present complex-real transition, and the complex energies lead to an unstably dynamical process of $E^{R}(t)$ while
the real energies result in a stable one \cite{Ginibre-MBL-entropy,NH_MBL_3}. Here in our model without time-reversal symmetry,
the energy spectra are all complex, but $E^{R}(t)$ presents two types of dynamical behaviors. It indicates that the stable
dynamical evolution of $E^{R}(t)$ is independent of time-reversal symmetry and real energy spectrum.

\section{Many-body localization transition}\label{S4}
\begin{figure}[htp]
	\centering
	\includegraphics[width=0.5\textwidth]{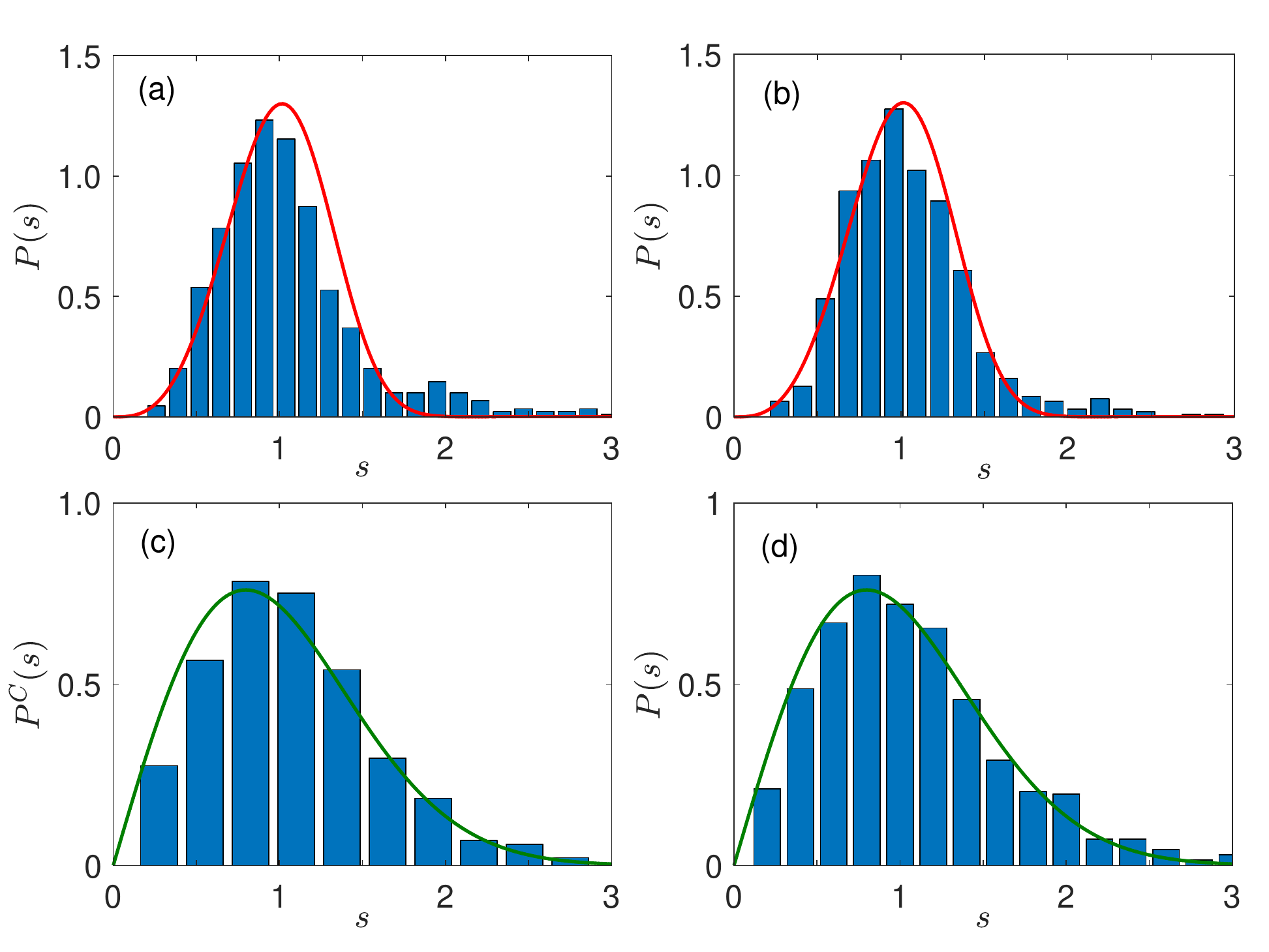}
	\caption{The nearest-level-spacing distribution of the unfolded energies for (a) $V=2J$,
	 (b) $V=3J$, (c) $V=5J$, and (d) $V=8J$. The red curve is the Ginibre distribution function $P^{C}_{Gin}$
	and the green curve is the Poisson distribution function $P^{C}_{Po}(s)=\pi s/2e^{-(\pi/4)s^2}$ on the
	complex energy plane. The energies used in the statistics are taken from the middle of the energy spectra with
	a proportion of $20\%$.
	Other parameters are {\color{red} $\gamma=0.1J$}, $L=14$ and $\varphi=\pi/3$.} \label{f4}
\end{figure}

Inspired by early estimate on the location of MBL point of Hermitian system \cite{Huse_MBL_AA} and
recent works on non-Hermitian MBL \cite{Ginibre-MBL-entropy,NH_MBL_2,NH_MBL_3,NH_MBL_4,Suthar_1},
we are aware of that the stable dynamical behavior of $E^{R}(t)$ in the system without time-reversal symmetry and
real energy spectrum may be caused by the MBL. We first study the nearest-level-spacing statistic to check
the mentioned point. If a non-Hermitian many-body system exists MBL transition, there are different spectral statistic laws
before and after MBL transition \cite{Ginibre-MBL-entropy,Ginibre-symmetry}.  For an energy $E_{\beta}$, the nearest-level spacing
is defined by the minimal distance in the complex energy plane, i.e., ${\rm min}_{\beta'}|E_{\beta}-E_{\beta'}|$. For the delocalized
phase, it is known that the statistical distribution obeys the Ginibre distribution $P^{C}_{Gin}(s)=cp(cs)$ \cite{Ginibre_para_1,Ginibre_para_2}, in which
\begin{equation*}
p(s)=\lim_{M\rightarrow \infty}\left[\prod^{M-1}_{m=1}e_{m}(s^2)e^{-s^2}\right]\sum^{M-1}_{m=1}\frac{2s^{2m+1}}{m!e_{m}(s^2)}
\end{equation*}
with $e_{m}(s^2)=\sum^{m}_{\ell=0}s^{2\ell}/\ell !$ and $c=\int^{\infty}_{0}sp(s)ds=1.1429\cdots$. In our numerical computing,
energies used in the statistics are taken from the middle of the energy spectra with a proportion of $20\%$. For parameters
$V$ taken before $V_c$, such as $V=2J$ and $V=3J$, the corresponding statistical distributions are shown in Figs.~\ref{f4}(a)
and \ref{f4}(b), respectively. The distributions match with the Ginibre distribution (the red curve shows).
For larger parameters exceeding $V_{c}$, such as $V=5J$ and $V=8J$, the corresponding statistical distributions are plotted
in Figs.~\ref{f4}(a) and \ref{f4}(b), respectively. Intuitively, the distributions are matched with the Poisson distribution $P^{C}_{Po}(s)$
on the complex plane \cite{Ginibre-MBL-entropy,NH_MBL_3,NH_MBL_4}, satisfying $P^{C}_{Po}(s)=\pi s/2e^{-(\pi/4)s^2}$.
According to the symmetry classes of non-Hermitian random matrices, here our model shall belong to the A class \cite{Ginibre-symmetry}.
Although the A class and the complex-conjugation symmetry (or say time-reversal symmetry) class \cite{Ginibre-MBL-entropy,NH_MBL_2,NH_MBL_3,NH_MBL_4}
are members of the Ginibre's symmetry classes \cite{Ginibre-symmetry}, the corresponding statistical characteristics are different.
To be specific, in our model, the spectral statistics present a transition from the Ginibre distribution to the complex
Poisson distribution (which has the same distribution function as the Wigner-Dyson distribution), whereas in the
complex-conjugation symmetry class, the spectral statistics display a transition from the Ginibre distribution to the
real Poisson distribution. In the complex-conjugation class, the statistic distribution transition is accompanied by the MBL transition.
The different statistical rules of energy spectrum suggest that the system has experienced the MBL, and also implies that the
MBL plays a leading role in maintaining the stable dynamic evolution of $E^{R}(t)$.

To further check the presence of MBL, we study the half-chain entanglement entropy and the IPR, which
are based on the many-body eigenstate. The half-chain entanglement entropy is defined as
\begin{equation}
S=-{\rm{Tr}}(\rho^{r}\ln\rho^{r}),
\end{equation}
where $\rho^{r}={\rm{Tr}}_{L/2}[\ket{\psi^{r}}\bra{\psi^{r}}]$ is the reduced density matrix with $\ket{\psi^{r}}$ the
right eigenstate. It is known that for the delocalized phase, entanglement entropy obeys the volume law and for the localized phase it obeys
the area law \cite{Ginibre-MBL-entropy}. Therefore, the entanglement entropy for the delocalized
phase is visibly larger than that of the localized phase \cite{Ginibre-MBL-entropy,NH_MBL_2,NH_MBL_3,NH_MBL_4}.
In view of this characteristic, entanglement entropy is usually used to distinguish the delocalized phase from the MBL one.

\begin{figure}[htp]
	\centering
	\includegraphics[width=0.5\textwidth]{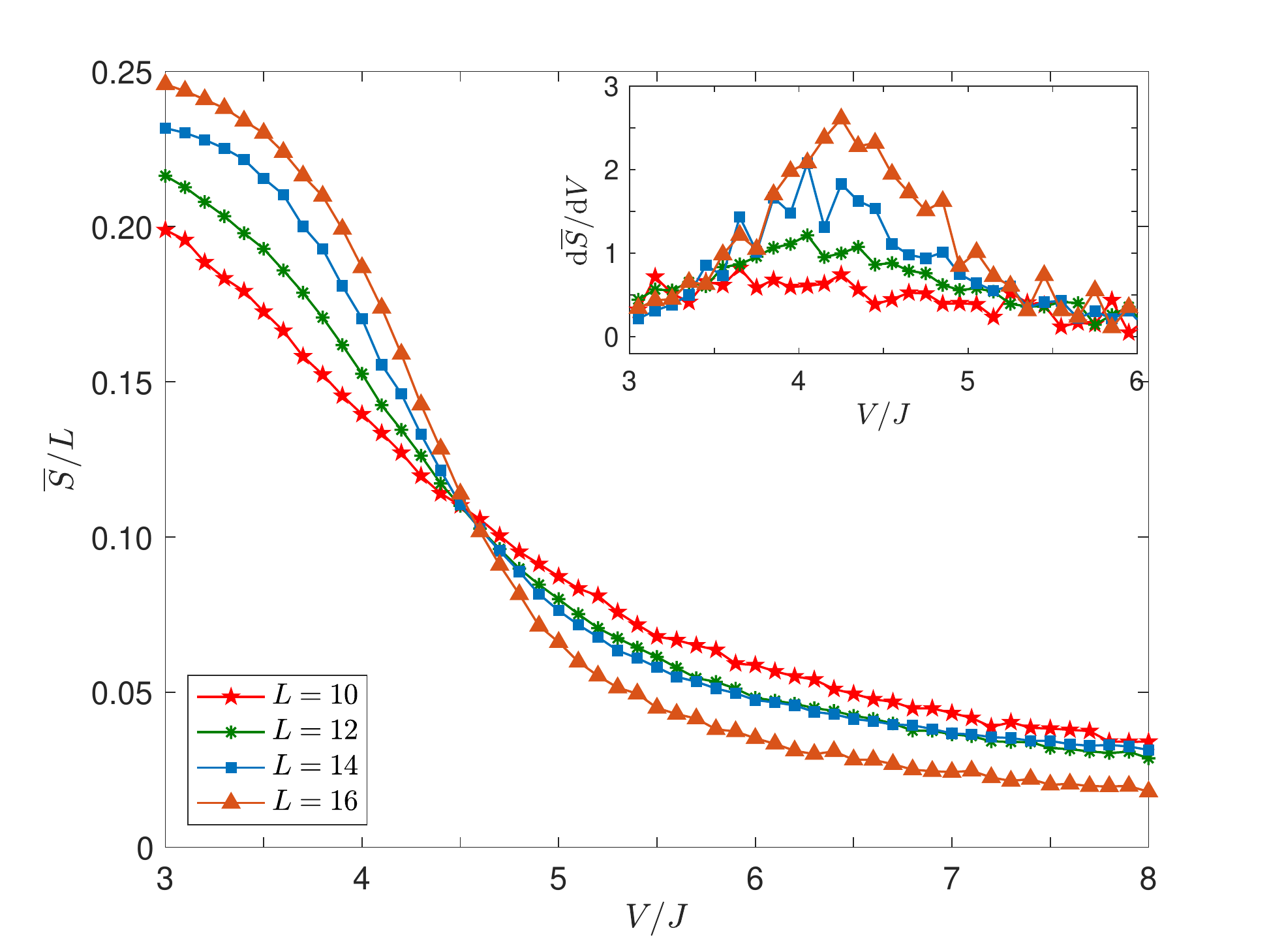}
	\caption{ Ensemble average of the half-chain entanglement entropy over system size $\overline{S}/L$ as a function
	of the on-site potential strength $V$. As the increase of $V$, $\overline{S}/L$ shows crossover at $V^{MBL}_{c}=4.5\pm0.1J$,
	signaling the appearance of the MBL transition. Here, the right eigenstates are taken from the middle of the energy spectra
	with a proportion of $4\%$. The inset shows the slope of $\overline{S}$. We take $500$ samples for $L=10,~12$, $200$ samples
	for $L=14$, and $96$ samples for $L=16$.  Other parameter is $\gamma=0.1J$.
	} \label{f5}
\end{figure}

Under different system sizes and $\gamma=0.1J$, we take the eigenstates from the
middle $4\%$ spectrum to calculate the corresponding entanglement entropy. The ensemble-averaged
entropy $\overline{S}$ over the system size $L$, i.e., $\overline{S}/L$ are plotted in Fig.~\ref{f5}.
Generally speaking, it is a common and effective method to determine the location of MBL transition
by the the crossing of entanglement entropy curves. But there are exceptions where the crossing shifts with system size,
so that the existence of the absolutely stable MBL is under debate \cite{Prosen,Sels,Luitz_and_Huse}. For the current studied model, the crossing
of entanglement entropy curves does not shift with system size. Therefore, it is feasible to determine the localization
of MBL transition from the crossing of the entanglement entropy curves.  As can be seen that the entropy
exhibits a system size independent crossover at the transition point $V^{\rm{MBL}}_{c}\simeq 4.5\pm 0.1J$.
We note that the MBL transition point of our system is less than that of the system with time reversal
symmetry \cite{NH_MBL_3}. It means that the absence of the time-reversal symmetry can suppress the delocalization, and cause the
MBL to occur earlier in the system. Besides, the entropy presents a transition from the volume law to the area law.
Before $V^{\rm{MBL}}_{c}$ the entropy is larger than that after $V^{\rm{MBL}}_{c}$. Meanwhile, the entropy shows an $L$-dependence.
When $V$ is less than $V^{\rm{MBL}}_{c}$, the entropy increases as $L$ increases, and when $V$ is larger than $V^{\rm{MBL}}_{c}$,
the entropy decreases as $L$ increases. Besides, we can employ the slope of $\overline{S}$ to estimate the MBL transition point.
Because $\overline{S}$ decays with the increase of $V$, the definition of the slope of $\overline{S}$, i.e.,
${\rm d} \overline{S}/{\rm d}V=\left[\overline{S(V)}-\overline{S(V+\delta V)}\right]/\delta V$ is beneficial for visualization.
In the calculation, we take $\delta V=0.1J$ and the slopes for different $L$ are plotted in the inset of Fig.~\ref{f5}.
It shows that the slopes peak near the MBL transition point, and the feature is more evident for larger system sizes.

\begin{figure}[htp]
	\centering
	\includegraphics[width=0.5\textwidth]{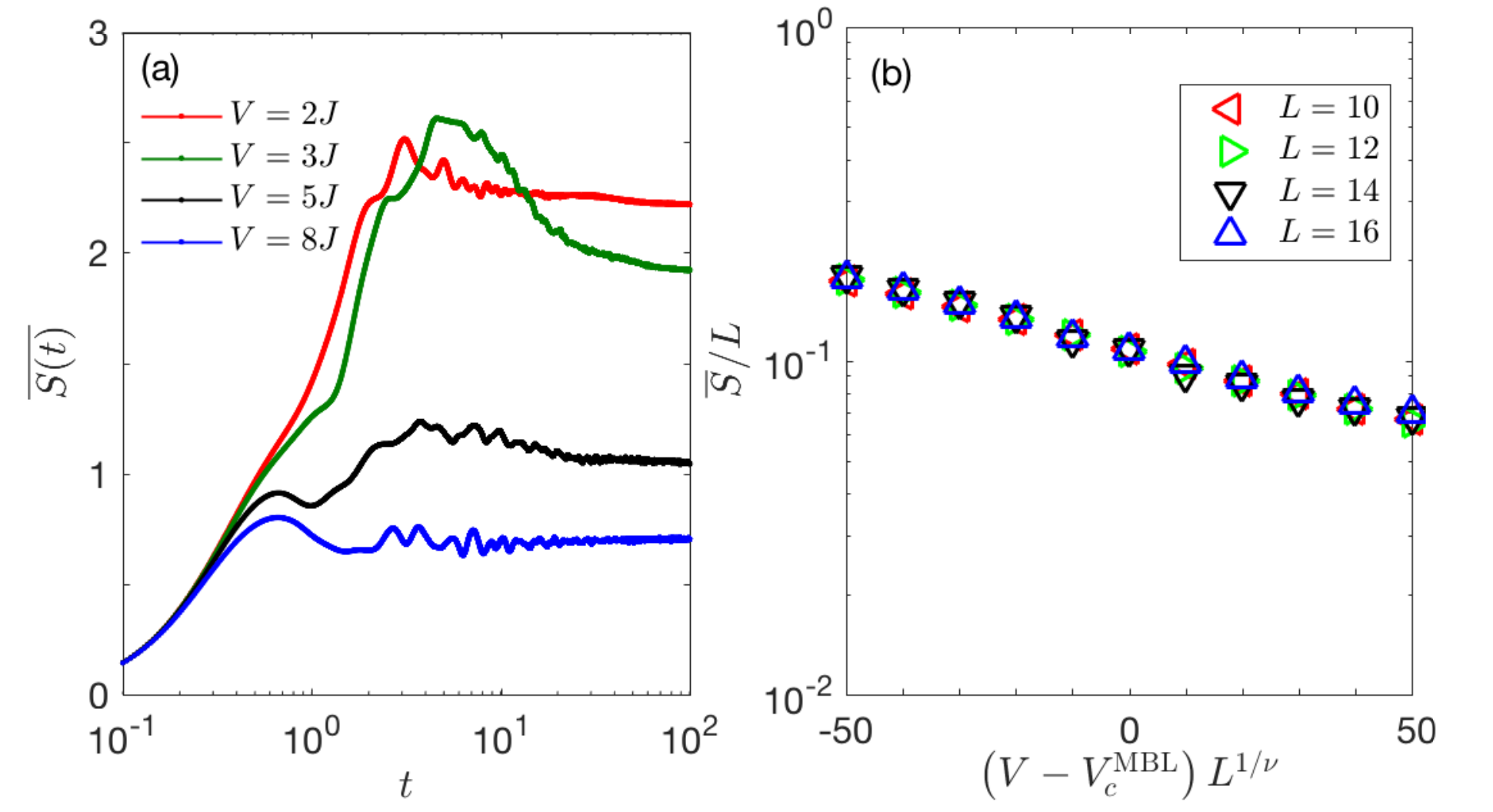}
	\caption{ (a) Ensemble average of the time-dependent half-chain entanglement entropy $\overline{S(t)}$ for various $V$.
	We take $\ket{\psi(t=0)}=\ket{101010\cdots}$ as the initial state, and use $L=14$ and $512$ samples in the calculations.
	(b) Finite size scaling collapse of the entropy as a function of $(V-V^{\rm MBL}_{c})L^{1/\nu}$, where we
	take $V^{\rm MBL}_{c})=4.55J$ and $\nu=0.6$. The associated eigenstates are taken from the middle $4\%$ spectrum.
	We take $512$ samples for $L=10$, $12$, and 256 samples for $L=14$, and $32$ samples for $L=16$.
	 Other parameter is $\gamma=0.1J$.
	}
	 \label{f6}
\end{figure}

The MBL transition can be characterized by the dynamical behaviors of the entanglement entropy as well. Still taking $\ket{\psi(t=0)}=\ket{101010\cdots}$
as the initial state, we calculate the ensemble average of the time-dependent entropy $\overline{S(t)}$ under the system size $L=14$,
and the results are shown in Fig.~\ref{f6}(a). It can be seen that when evolution time $\tau$ is short, $\overline{S}(t)$ for different $V$
synchronously grow with $\tau$, and soon enter their own evolutionary trajectories, and finally tend to steady values. For parameters $V=2J$
and $V=3J$ chosen in the delocalized phase, we can see that $\overline{S(t)}$ is larger than that chosen from the MBL phase ($V=5J$ and $V=8J$ cases).
Moreover, the entanglement entropies of the steady states gradually decreases with the increase of $V$, and this variation tendency is
consistent with that of the static entanglement entropy. From the entanglement entropy and its dynamical behaviors, we know that there
actually exists a delocalization to MBL transition behind the complex energy, two types of different dynamical phenomena of $E^{R}(t)$,
and the transition from Ginibre's distribution to complex Poisson distribution. To reveal the critical exponent of the
MBL transition, we will study the scaling behavior of the entanglement entropy near the MBL transition point. For a finite size system,
the entanglement entropy around $V^{\rm{MBL}}_{c}$ satisfies the following scaling behavior \cite{Ginibre-MBL-entropy}
\begin{equation}
S/L=f\left[(V-V^{\rm{MBL}}_{c})L^{1/\nu}\right],
\end{equation}
where $\nu$ is the critical exponent and $f(x)$ is the scaling function. With $V^{\rm{MBL}}_{c}=4.55J$ and different system size,
the corresponding ensemble averages of the entropy over the system size $\overline{S}/L$ are plotted in Fig~\ref{f6}(b). It shows
that $\overline{S}/L$ for different $L$ collapse onto a single curve with $\nu\simeq 0.6$, different from the scaling exponent $\nu=1.3$ in
the complex-conjugation symmetry class \cite{Ginibre-MBL-entropy,NH_MBL_4}, $\nu=1.8$ in the transposition symmetry
class \cite{Ginibre-MBL-entropy}, and $\nu=1$ and $\nu=1.5$, corresponding to the cases with
non-reciproca hopping parameters $g=0.3$ and $g=0.6$, respectively \cite{NH_MBL_4}.

\begin{figure}[htp]
	\centering
	\includegraphics[width=0.5\textwidth]{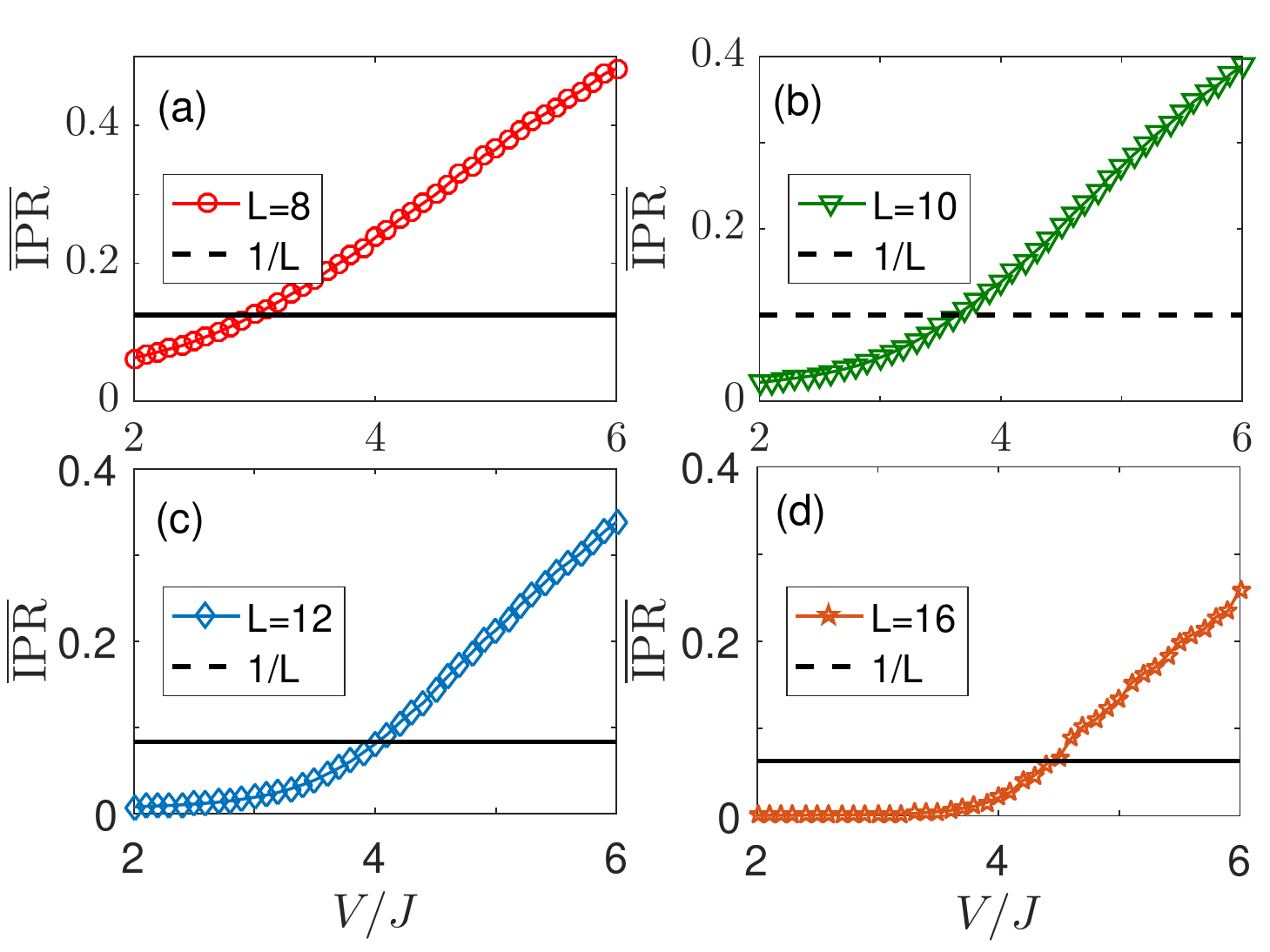}
	\caption{ Ensemble averaged IPR, i.e. $\overline{\rm IPR}$ as a function of $V$.
	We take $1000$ samples for (a) $L=8$, (b) $10$, (c) $12$, and $100$ samples for (d) $L=16$.
	 Other parameter is $\gamma=0.1J$.
	}
	 \label{f7}
\end{figure}

Theoretically, IPR plays a role of indicator to reveal the properties of the eigenstates in the noninteracting systems.
The IPR of extended eigenstates scales like $1/L$ and approaches finite constant for the localized states \cite{Xiexincheng}.
We try to employ this single-particle feature (i.e., ${\rm IPR} \propto 1/L$) to estimate the MBL transition point (marked by $V^{\rm IPR}_{c}$).
The IPR of a many-body eigenstate reads
\begin{equation}
{\rm IPR}=\sum^{Dim}_{k}|\psi(k)|^4,
\end{equation}
where $\psi(k)$ is the amplitude of the eigenstate $\ket{\psi}$ in the Fock basis $\{\ket{k}\}$ with $\psi(k)=\braket{k|\psi}$,
and $Dim$ is the size of the Hilbert space. Under different system sizes, we can obtain the corresponding ensemble averaged
IPR, i.e., $\overline{\rm IPR}$ as a function of $V$, and the $V$ solution to ${\rm IPR}=1/L$ is the estimated $V^{\rm IPR}_{c}$.
Taking $L=10$, $12$, $14$, and $16$, $\overline{\rm IPR}$ as a function of $V$ are plotted
in FIgs.~\ref{f7}(a)-\ref{f7}(d). For various system sizes, $\overline{\rm IPR}$ increases from a near-zero constant to a finite constant
as $V$ increases, presenting the delocalization to MBL transition. Meanwhile, there is always a solution $V^{\rm IPR}_{c}$
to $\overline{\rm IPR}=1/L$. $V^{\rm IPR}_{c}$ presents a dependence on the finite-size effect. When $L$ is small, as
shown in Fig.~\ref{f7}(a), \ref{f7}(b), and \ref{f7}(c), $V^{\rm IPR}_{c}$ deviates from $V^{\rm MBL}_{c}$. Until a larger system
size $L=16$, the estimated MBL transition point $V^{\rm IPR}_{c}=4.4J\sim 4.5J$ is much closer to $V^{\rm MBL}_{c}$. The slight
deviation can be attributed to the finite size effect. It shows that the solution to ${\rm IPR}=1/L$ can be used to estimate the
MBL transition point.

In experiments, the dynamics of the density imbalance (denoted by $I(t)$) is an observable measurement to detect the many-body localization \cite{atom-particle-imb-1,atom-particle-imb-2,atom-particle-imb-3,atom-particle-imb-4,atom-particle-imb-5,
atom-particle-imb-6,atom-particle-imb-7,processor_4,processor_5},
which is defined as
\begin{equation}
I(t)=\frac{n_{o}(t)-n_{e}(t)}{N},
\end{equation}
where $n_{o}(t)$ and $n_{e}(t)$ are the time-dependent densities (particle populations) at odd and even sites, respectively.
It was studied that in the long-time evolution limit, for the delocalization phase, $I(t)$ is stable at a finite value, implying that some initial
information is preserved, while for the many-body localized phase, $I(t)$ approaches zero, implying that the initial formation is
lost \cite{atom-particle-imb-1,atom-particle-imb-2,atom-particle-imb-3,atom-particle-imb-4,atom-particle-imb-5,
atom-particle-imb-6,atom-particle-imb-7,processor_4,processor_5}. Next, we employ the quantity $I(t)$ to detect the non-Hermitian many-body localization.

\begin{figure}[htp]
	\centering
	\includegraphics[width=0.5\textwidth]{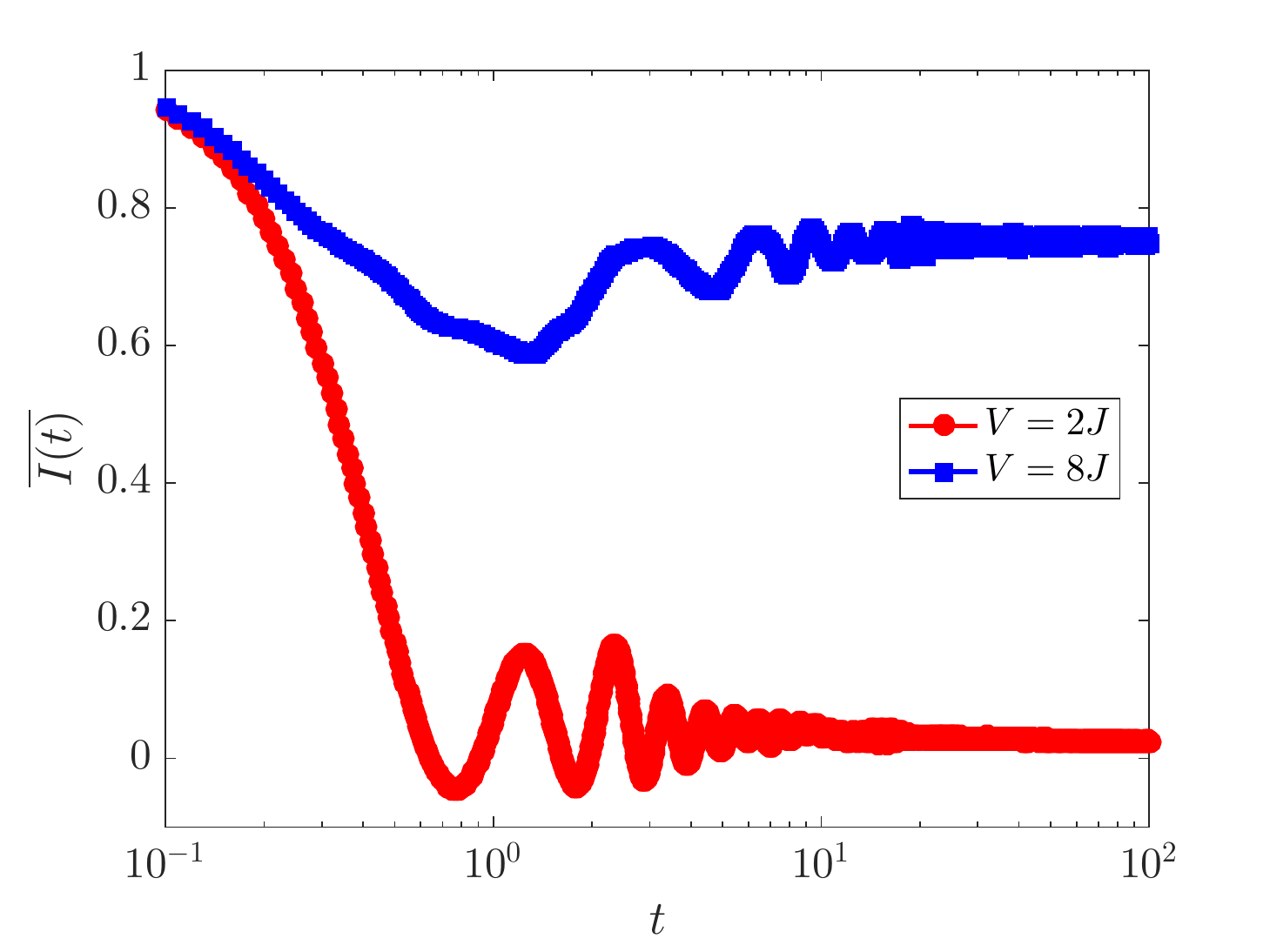}
	\caption{ Ensemble averaged density imbalance $\overline{I(t)}$ with different $V$.
	We take $512$ samples, {\color{red} $\gamma=0.1J$},  and $L=12$.}
	 \label{f8}
\end{figure}

Taking the initial state $\ket{\psi(t=0)}=\ket{101010\cdots}$, $\gamma=0.1J$, and $L=12$, the ensemble
averaged density imbalance $\overline{I(t)}$ as a function of the evolution time with different $V$ are plotted in Fig.~\ref{f8}. Intuitively, for $V=2J$, $\overline{I(t)}$ is
a finite value in the long-time evolution limit, which implies that the system is in the many-body delocalized phase. On the contrary, for $V=8J$,
$\overline{I(t)}$ approaches to zero in the long-time evolution limit, denoting that the system is in the many-body localized phase.
 Through the above researches on energy spectrum and its statistical law, half-chain entanglement entropy, IPR, and
density imbalance, we have clearly understood that the stable dynamic evolution of $E^{R}(t)$ attributed to the MBL of the system,
and has nothing to do with the real or complex energy spectrum.

\section{ topological transition}\label{S5}
\begin{figure}[!htp]
	\centering
	\includegraphics[width=0.5\textwidth]{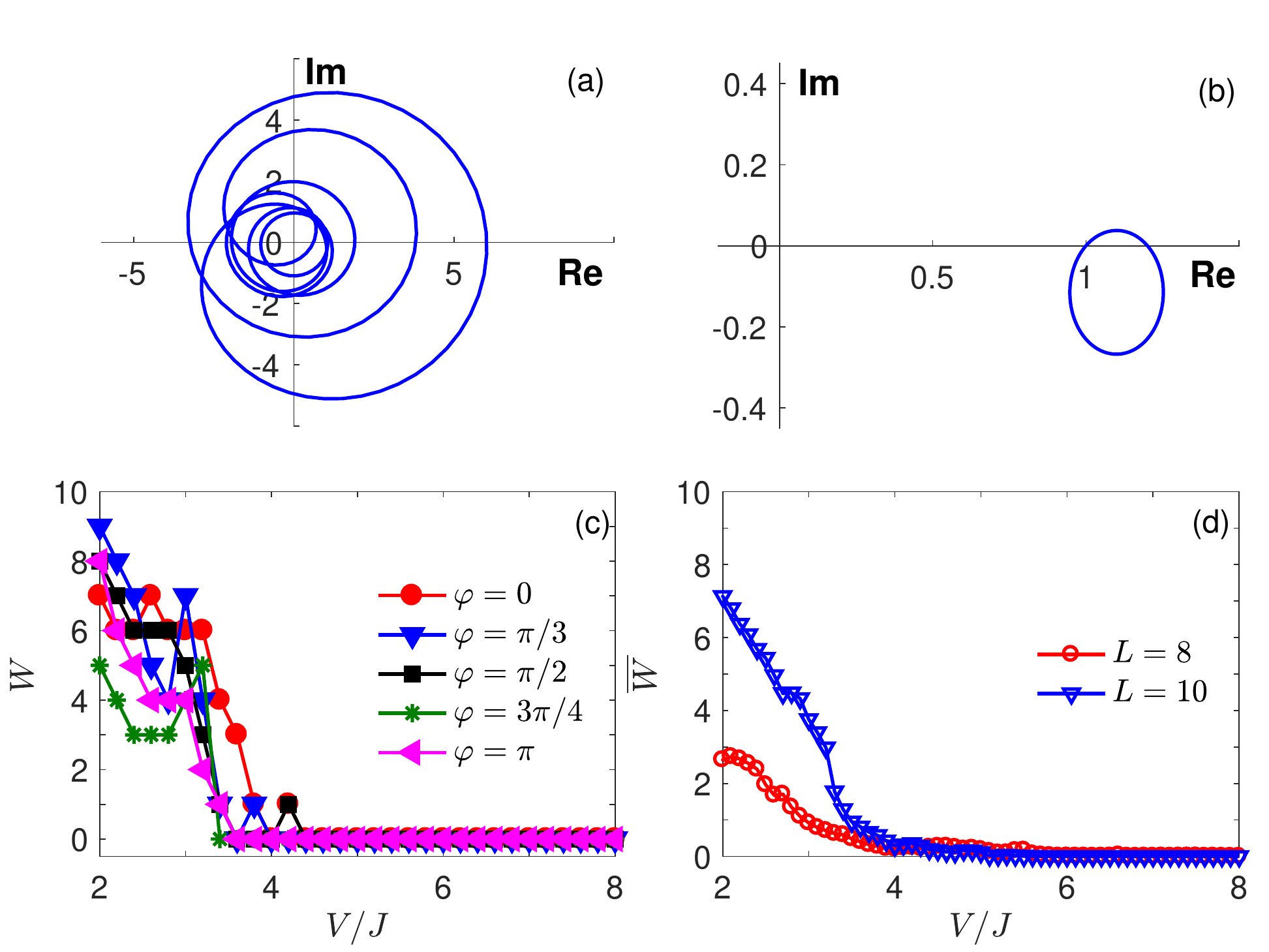}
	\caption{Trajectory of $\det H(\phi)/\det H(0)$ in the complex plane for (a) $V=3J$ and (b) $V=5J$,
	where the parameters chosen are $L=10$ and $\varphi=\pi/3$.
	(c) The $\varphi$ dependence of $W$ under the system size $L=10$.
	(d) The average of $W$ after $1000$ ensemble averages.  Other parameter is $\gamma=0.1J$.} \label{f3}
\end{figure}

We note that the complex-real transition and the non-Hermitian MBL transition is found to be accompanied by the topological transition
in the time-reversal symmetric case \cite{NH_MBL_3}. Meanwhile, there is no topological transition uncovered in a many-body complex
energy spectrum case \cite{NH_MBL_4}. It drives us to investigate whether there is topological transition in the current studied system
with complex energy spectrum, and whether the topological transition is accompanied by the MBL transition. To answer these questions,
we calculate the winding number \cite{NH_winding_1,NH_winding_2}. We perform a gauge transformation on the creation and annihilation
operators, i.e., $\hat{c}_{j}\rightarrow e^{i\frac{\phi}{L}j}\hat{c}_{j}$
and $\hat{c}^{\dag}_{j}\rightarrow e^{-i\frac{\phi}{L}j}\hat{c}^{\dag}_{j}$. Then the $\phi$-dependent Hamiltonian
$\hat{H}(\phi)$ reads
\begin{equation}
\begin{aligned}
\hat{H}(\phi)&=\sum^{L}_{j=1}\left[-J\left(e^{-g}e^{-i\frac{\phi}{L}j}\hat{c}^{\dag}_{j+1}\hat{c}_{j}+e^{g}e^{i\frac{\phi}{L}j}\hat{c}^{\dag}_{j}\hat{c}_{j+1}\right) \right.\\
&\left.+U\hat{n}_{j}\hat{n}_{j+1}+V_j\hat{n}_{j}
\right],
\end{aligned}
\end{equation}
and accordingly the winding number is defined as
\begin{equation}
W=\int^{2\pi}_{0}\frac{d\phi}{2\pi i}\partial_{\phi}\ln\det\{H(\phi)-E_{b}\},
\end{equation}
where $E_{b}$ is just the base energy. In the calculation, $E_b$ is taken at $E_b=0$. We emphasize that here the topology
is reflected in the trajectory of $\det H(\phi)/\det H(0)$ based on $H(\phi)$, and is quantized by the
winding number \cite{NH_winding_1,NH_winding_2}. When the closed trajectory of $\det H(\phi)/\det H(0)$ encircles $E_b$ once, the winding number
accumulates by $1$. If the closed trajectory does not encircle $E_b$ or there is no closed trajectory, the winding number
is zero \cite{NH_MBL_3}. With system size $L=10$ and a single phase $\varphi=\pi/3$, the trajectories of $\det H(\phi)/\det H(0)$ for different
$V$ are plotted in Figs.~\ref{f3}(a) and \ref{f3}(b), respectively. As shown in Fig.~\ref{f3}(a), the trajectory of $\det H(\phi)/\det H(0)$
encircles the base energy seven times, leading to the winding number $W=7$. In Fig.~\ref{f3}(b), it is seen that although there is a
closed trajectory in the complex plane, the trajectory does not encircle the base energy, resulting in the winding number $W=0$.
The results show that there actually exists topological transition even if the energy spectra are all complex. Meanwhile, 
the results also show that the topological transition is not intrinsically related to the real-complex transition of the energy spectrum. 
It is noted that there is no tight connection between the MBL transition and the real-complex transition of the energy spectrum, 
we therefore believe that the MBL transition is the main cause of topological transition.

Note that here the frequency of the on-site potential is incommensurate which leads to a $\varphi$-dependent energy spectrum.
Therefore, for different $\varphi$, the corresponding winding number at a specific $V$ will be different. In Fig.~\ref{f3}(c), $W$ as a
function of $V$ with $\varphi=0$ (red), $\varphi=\pi/3$ (blue), $\varphi=\pi/2$ (black), $\varphi=3\pi/4$ (green), and
$\varphi=\pi$ (magenta) are plotted. Here, the system size is taken at $L=10$. From this diagram, we can see that although the
winding number $W$ presents a $\varphi$ dependence, there exists a feature that a critical potential strength $V_c$ can
divide the system into two parts. Before $V_c$, the system is non-trivial with non-zero $W$, and after $V_c$, the system is trivial with $W=0$.
To extract the critical parameter $V_{c}$, we perform the finite size analysis. Figure \ref{f3}(d) shows the
averages of $W$ as the function of $V$ with two different system sizes. Here we have averaged $1000$ ensembles. We can see
that the two curves intersect about $V=4.4\pm 0.1J$, which implies that the transition point of topological transition is
about $V_{c}=4.4\pm0.1J$. The results show that the topological transition actually exists in this current studied
system without real energy spectrum and time-reversal symmetry. By comparing the transition points of MBL and topological transition,
we know that their transition points overlap, indicating that the topological transition occurs synchronously with the MBL transition.

\begin{figure}[!htp]
	\centering
	\includegraphics[width=0.5\textwidth]{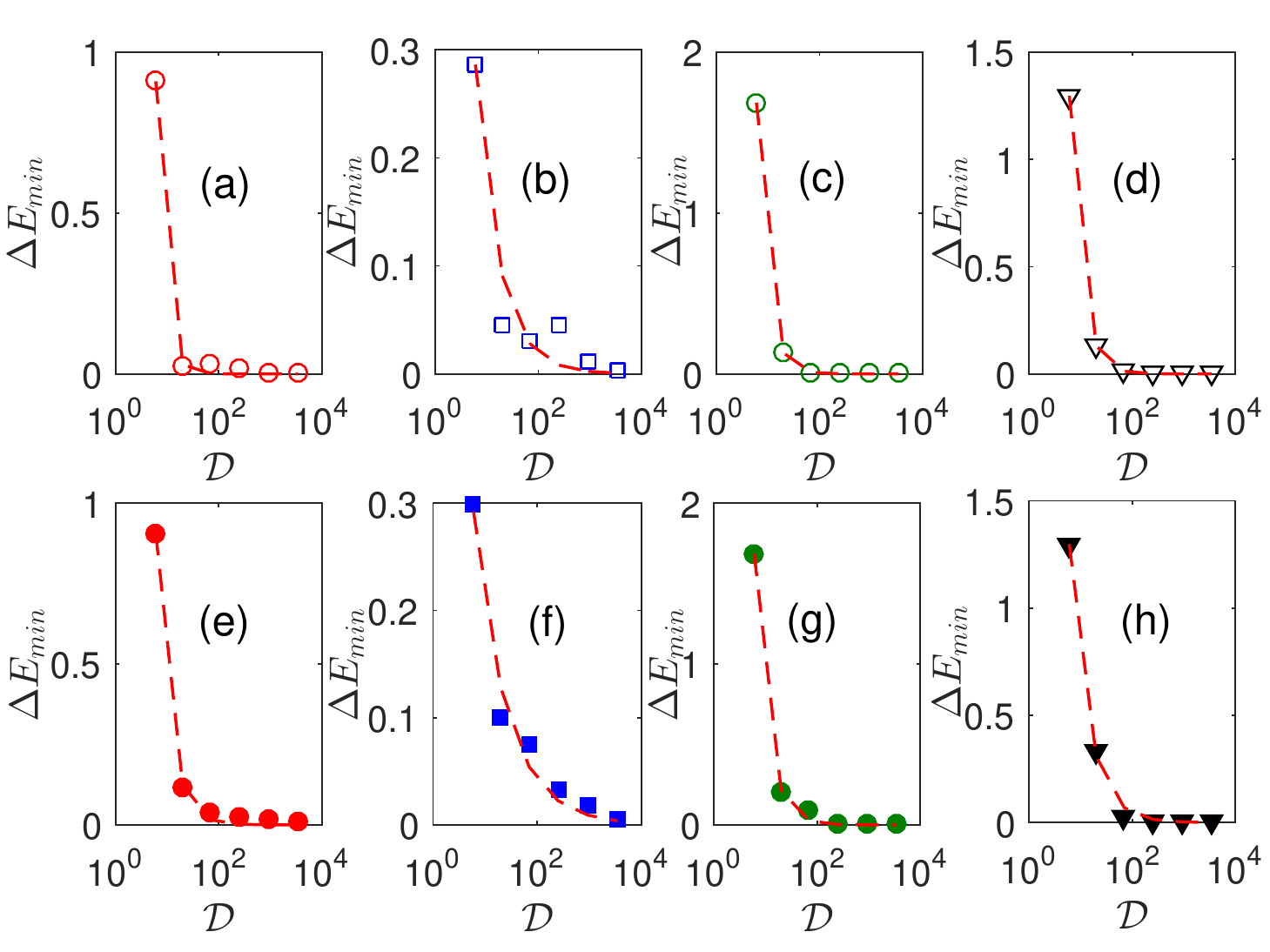}
	\caption{Minimal energy gap $\Delta{E}_{gap}$ as a function of the size of system subspace $\mathcal{D}$.
(a) $V=2J$, $\varphi=0$, and $\gamma=0$. (b) $V=2J$, $\varphi=\pi/3$, and $\gamma=0$.
(c) $V=8J$, $\varphi=0$, and $\gamma=0$. (d) $V=8J$, $\varphi=\pi/3$, and $\gamma=0$.
(e) $V=2J$, $\varphi=0$, and $\gamma=0.1$. (f) $V=2J$, $\varphi=\pi/3$, and $\gamma=0.1$.
(g) $V=8J$, $\varphi=0$, and $\gamma=0.1$. (h) $V=8J$, $\varphi=\pi/3$, and $\gamma=0.1$.
The discrete data are the calculated minimal energy gap and the red dashed lines are the corresponding fitting curves.} \label{f3_1}
\end{figure}

In addition, it was studied that the exceptional point \cite{Heiss_1990,Heiss_2005,Heiss_2012,PRE2018} could lead to
the non-trivial topology \cite{Kawabata_PRB,Kawabata_PRL,PRR_Luitz}. Therefore, we will investigate the connection
between the uncovered topological phase transition and the exceptional points. In order to determine whether there are
exceptional points in the system, we define the minimal energy gap
 $\Delta{E}_{gap}$ as $\Delta{E}_{gap}={\rm min}_{\beta,\beta'}|E_{\beta}-E_{\beta'}|$, where $E_{\beta}$ and $E_{\beta'}$ are two different energies.
At first, we study the $\gamma=0$ case. With different $V$ and $\varphi$, the corresponding $\Delta{E}_{gap}$
(shown with the discrete data points) are plotted in Figs.~\ref{f3_1}(a), \ref{f3_1}(b), \ref{f3_1}(c), and \ref{f3_1}(d) respectively.
The red dashed lines are the fitting curves, satisfying $f(\mathcal{D})=a\mathcal{D}^{b}$ (where $a$ and $b$ are the fitting parameters
with $a$ being a positive number and $b$ being a negative number, presented in Table \ref{t1},
and $\mathcal{D}$ is the size of the system subspace). Here, we choose the inverse power of the system subspace
as the fitting function because the system subspace grows exponentially with the system size \cite{Huse_MBL_AA}. The fitting results show
that when the system subspace tends to infinite, $\Delta{E}_{min}$ under different $V$ all approach to zero, indicating that
there exist exceptional points regardless of whether the potential strength $V$ is small or large. Meanwhile,
the decreasing of $\Delta{E}_{min}$ as an inverse power of $\mathcal{D}$ indicates that the system subspace is
exponentially dependent of the number of particles.
We note that in the $\gamma=0$ case, $V=2J$ corresponds to non-zero winding number, and $V=8J$ corresponds
to zero winding number \cite{NH_MBL_3}. It means that the topological transition followed by complex-real energy transition
in Ref. \cite{NH_MBL_3} is not caused by the exceptional points. Next, we investigate whether the topological phase transition
in our complex energy model is related to exceptional points. Taking different $V$, the
corresponding $\Delta{E}_{gap}$ (shown with discrete data points)
as a function of the system size are plotted in Figs.~\ref{f3_1}(e), \ref{f3_1}(f), \ref{f3_1}(g), and \ref{f3_1}(h), respectively.
The fitting results (see Table \ref{t1} for details) show that $\Delta{E}_{gap}$ decays with system size in
a power law form at chosen $V$, and approaches to zero at larger system size. It indicates that in our complex energy model,
there are exceptional points as well, regardless of whether the potential strength is small or large. Noting that
when $V$ is changed from $V=2J$ to $V=8J$, the system undergoes a topological transition from non-zero $W$ to $W=0$.
As a result, similar to that in the system with complex-real energy transition, the non-trivial topology
in our complex energy model is independent of the exceptional point and originates from the non-trivial trajectory
of $\det H(\phi)/\det H(0)$ surrounding the base energy.

\begin{widetext}
\begin{table}[H]
\centering
\begin{tabular}{|c|c|c|c|c|c|c|c|c|}
 & Fig.\ref{f3_1}(a)& Fig.\ref{f3_1}(b) & Fig.\ref{f3_1}(c) & Fig.\ref{f3_1}(d) & Fig.\ref{f3_1}(e) & Fig.\ref{f3_1}(f) & Fig.\ref{f3_1}(g) & Fig.\ref{f3_1}(h) \\ \hline
a & $154.20$ & 1.582 & $75.92$ & 39.64 & 17.58& 1.00& 36.03& 11.03\\ \hline
b & -2.865 & -0.954 & -2.125 & -1.909 & -1.654 &-0.687 &-1.709 & -1.194 \\ \hline
\end{tabular}
\caption{The fitting parameters of the fitting curves presented in Fig.~\ref{f3_1}.}\label{t1}
\end{table}
\end{widetext}

Up to now, we have known that the topology origin of our complex energy system and
that the counterintuitive dynamical behaviors are caused by the MBL not the real energy spectrum.
Meanwhile, the topological transition is found to be simultaneous with the MBL transition, and the stability
of the time evolution of $E^{R}(t)$ can be predicted by the winding number which is defined in the complex
plane after a gauge transformation. In the following, we introduce how to realize the Hamiltonian presented in Eq. (\ref{Ham}).
The dynamical process of the density matrix $\rho$ for an open system is governed by a Lindblad master equation \cite{Lindblad}:
\begin{equation}
\dot{\rho}_{t}=-i\left[\mathcal{H},\rho\right]+\sum_{j}\mathcal{D}\left[L_{j}\right]\rho,
\end{equation}
where $\mathcal{H}$ is a Hermitian Hamiltonian, which is just the $H$ when
$g=\gamma \equiv 0$, $\mathcal{D}\left[L_{j}\right]=L_{j}\rho L^{\dag}_{j}-\{ L^{\dag}_{j}L_{j},\rho\}/2$,
and $L_{j}$ is the Lindblad dissipator describing the quantum jump between the system and the environment.
Under the postselection \cite{NH_winding_1} or no-jump condition \cite{time_evolution}, the Lindblad dynamical evolution can be governed by a non-Hermitian
effective Hamiltonian $H_{\rm eff}$, which is expressed as
\begin{equation}
H_{\mathrm{eff}}=\mathcal{H}-\frac{i}{2}\sum_{j}\ L_{j}^\dag L_{j}.
\end{equation}

We note that only considering the one-body loss can only realize the non-reciprocal hoppings, but can not 
realize the odd (even)-site gain (loss) \cite{NH_winding_1}. To achieve both, we shall consider local (or say, site-dependent) one-body 
loss $L^{loss}_{j}$ and gain $L^{gain}_{j}$, which are denoted as
\begin{equation}
\begin{aligned}
&L^{loss}_{j}=\sqrt{\kappa_{j}}\hat{c}_{j}\pm i \sqrt{\kappa_{j+1}}\hat{c}_{j+1}, \\
&L^{gain}_{j}=\sqrt{\beta_{j}}\hat{c}^{\dag}_{j}\mp i \sqrt{\beta_{j+1}}\hat{c}^{\dag}_{j+1},
\end{aligned}
\end{equation}
where $\kappa_{j}$ and $\beta_{j}$ are the strengths of the local one-body loss
and gain, respectively. This means that there are both gain and loss of particles at 
two nearest adjacent lattice sites of the system. Therefore, the summation in $H_{\rm eff}$ shall 
extend over all lattice sites and the dissipators (i.e., including the loss and gain).

Employing the commutation relation of the hard-core bosons $\left[\hat{c}_{j}, \hat{c}^{\dag}_{k}\right]=\delta_{jk}\left(1-2\hat{c}^{\dag}_{j}\hat{c}_{j}\right)$,
we arrive at $H_{\rm eff}$ as
\begin{equation}
\begin{aligned}
H_{\rm eff}&=\sum_{j}\left(J^{R}_{j} \hat{c}^{\dag}_{j+1}\hat{c}_{j}+J^{L}_{j} \hat{c}^{\dag}_{j} \hat{c}_{j+1}\right) +\sum_{j} V_{j} \hat{n}_{j} \\
& + \sum_{j}U\hat{n}_{j}\hat{n}_{j+1}+i\sum_{j}\gamma_{j}\hat{n}_{j}-i\sum_{j}\beta_{j},
\end{aligned}
\end{equation}
where
\begin{equation}
\begin{aligned}
J^{R}_{j}&=-J\mp\frac{\sqrt{\kappa_{j+1}}\cdot \sqrt{\kappa_{j}}}{2} \mp \frac{\sqrt{\beta_{j+1}}\cdot \sqrt{\beta_{j}}}{2}, \\
J^{L}_{j}&=-J\pm\frac{\sqrt{\kappa_{j+1}}\cdot \sqrt{\kappa_{j}}}{2} \pm \frac{\sqrt{\beta_{j+1}}\cdot \sqrt{\beta_{j}}}{2}, \\
\gamma_{j}&=\beta_{j}-\kappa_{j}.
\end{aligned}
\end{equation}
The last term $-i\sum_{j}\beta_{j }$ in $H_{\rm eff}$ denotes a background loss. If the strengths of site-dependent loss and gain are
staggered arranged, and satisfy the necessary condition, i.e., $|\beta_{j}-\kappa_{j}|=\gamma$, we can realize the Hamiltonian in Eq. (\ref{Ham}).
Therefore, we believe that with the help of current experimental techniques, the non-Hermitian MBL transition
and topological transition we have studied can be observed experimentally.

\section{Summary}\label{S6}
In the main text, we have investigated a non-Hermitian quasi-disordered many-body system that lacks real energy spectra.
Through analysis of spectral statistics, entanglement entropy, inverse participation ratio, and winding number, we have discovered that the many-body localization transition, the spectral statistics transition, and the topological transition
driven by quasi-disorder occur simultaneously. In the many-body delocalized phase, the energy spectra obey the Ginibre
distribution and the entanglement entropy obeys the volume law. Meanwhile, the inverse participation ratio approaches zero,
accompanied by a non-zero winding number in the delocalized phase. In contrast, in the many-body localized phase, the energy spectra obey
the complex Poisson distribution, and the entanglement entropy obeys the area law, accompanied by the finite inverse participation
ratio and zero winding number. By analyzing the non-Hermitian energy gap, we find that the energy gap  decays exponentially
with the system size, signaling the existence of the exceptional point. However, the non-Hermitian many-body
topology is independent of the exceptional point based on our findings that the exceptional point exists in both the non-zero winding
number and zero winding number phase regions, regardless of the presence or absence of real-complex transition. We argue
that this non-Hermitian many-body topology originates from the non-trivial trajectory of ${\rm det}H(\phi)/{\rm det}H(0)$ surrounding the
base energy. In addition, the critical exponent of the MBL transition is obtained. Furthermore, this many-body localization transition can be experimentally
observed by measuring the density imbalance \cite{atom-particle-imb-1,atom-particle-imb-2,atom-particle-imb-3,atom-particle-imb-4,atom-particle-imb-5,
atom-particle-imb-6,atom-particle-imb-7,processor_4,processor_5}. We have theoretically calculated the dynamics of density imbalance
and find that under the long-time evolution limit, the delocalization phase corresponds to near-zero density imbalance, whereas the
localized phase corresponds to finite density imbalance. It is noteworthy that despite the absence of real energy in the
many-body localization transition phase, the real part of the complex energy exhibited relatively stable dynamical behavior, indicating that the many-body
localization transition plays a critical role in maintaining this stability. We demonstrate that the Hamiltonian of the studied model can
be realized by the Lindblad dynamical evolution under postselection or no-jump condition. Recently, two works have studied non-Hermitian,
disorder-free many-body systems that exhibit complex-to-real energy transitions \cite{NH_Stark_MBL_1,NH_Stark_MBL_2}.
It would be interesting to investigate whether the many-body localization transition or the real energy is the crucial factor
in maintaining the stable dynamical behavior of the real part of the energy in such disorder-free systems.

We acknowledge support from NSFC under Grants No. 11835011 and No. 12174346.

\section*{References}

\bibliography{ref}

\end{document}